# Long-range phase coherence and phase patterns in hybrid Josephson junction arrays

Xi Wang[1], Asbjørn C. C. Drachmann[2], Candice Thomas[3], Michael J. Manfra[3-6], Nandini Trivedi[7], Charles M. Marcus[2,8], Saulius Vaitiekėnas[2,*], Beena Kalisky[1,*]

1. Department of Physics and Institute of Nanotechnology and Advanced Materials, Bar-Ilan University, Ramat Gan 5290002, Israel
2. Center for Quantum Devices, Niels Bohr Institute, University of Copenhagen, Copenhagen 2100, Denmark
3. Department of Physics and Astronomy, Purdue University, West Lafayette, Indiana 47907, USA
4. Purdue Quantum Science and Engineering Institute (PQSEI), Purdue University, West Lafayette, Indiana 47907, USA
5. Elmore Family School of Electrical and Computer Engineering, Purdue University, West Lafayette, Indiana 47907, USA
6. School of Materials Engineering, Purdue University, West Lafayette, Indiana 47907, USA
7. Department of Physics, Ohio State University, Columbus, Ohio 43210, USA
8. Materials Science and Engineering, and Department of Physics, University of Washington, Seattle, Washington 98195, USA

saulius@nbi.ku.dk, beena@biu.ac.il

**The coherence of superconductivity and its suppression near a quantum phase transition is governed by the interplay between local pairing and macroscopic phase coherence. Using scanning SQUID, we image the local susceptibility in a hybrid Josephson junction array. On a square lattice of narrow islands, we simultaneously access both the amplitude and spatial phase structure of sensitive superconducting states. We observe periodic phase patterns at commensurate magnetic fillings. At zero field the long-range phase coherence is strongest. At a finite field, smaller than one percent of flux quantum per unit cell, the system fragments into large regions of constant superconducting phase, as a function of the applied field. Our results provide the first direct measurement of long-range phase coherence in a Josephson junction array.**

A superconductor is described by a complex order parameter $\Delta = |\Delta|e^{i\phi}$ with an amplitude $|\Delta|$ and a phase $\phi$ [1]. Perturbations such as temperature, disorder, magnetic fields, or carrier density tuning can drive the system out of the superconducting (SC) state, either by breaking Cooper pairs or by disrupting their phase coherence [2]. Phase variations can manifest as circulating vortices or collective excitations [3], reflecting the rich spatial structure of the SC state.

Josephson junctions provide a natural platform for controlling the relative phase between SC regions, making them ideal to directly probe the dynamics of the relative phase degree of freedom [4,5]. Josephson junction arrays (JJAs) extend this concept to large ensembles of coupled elements, enabling systematic exploration of long-range phase coherence and collective phenomena [6,7]. The distribution of Josephson coupling $E_J$ and charging energy $E_c$ across an array governs the behavior of JJA [7–9]. Precise knowledge of these parameters is crucial both for characterizing individual devices and for using JJAs as a platform to study quantum phase transitions under tunable conditions, such as electrostatic gate voltages and magnetic fields.

In homogeneously disordered SC films, the superconductor-to-insulator transition (SIT) occurs via phase disordering. When $E_C$ dominates, Cooper pairs remain intact within SC puddles on a scale of the coherence length $\xi$, while long-range phase coherence is lost as the superfluid stiffness vanishes [10–16]. In contrast, when $E_J$ dominates, these puddles can establish phase coherence over a much larger scale $\xi_C \gg \xi$, resulting in macroscopic superconductivity. JJAs enable the study of phase coherence without the complications introduced by intrinsic disorder [17–19]. Hybrid JJAs further extend this capability by incorporating tunable materials [17–22], enabling in situ control of the ratio $E_C/E_J$ through adjusting the coupling between SC islands via etching, annealing, or electrostatic gating.

The spatial variation of $|\Delta|$ has been mapped by scanning tunneling microscopy (STM) with a resolution on the scale of the coherence length $\xi$ within individual domains of disordered and granular superconductors [23–25]. However, it is difficult to probe the relative phase ϕ between grains and over large $\xi_C \gg \xi$ scale. The phase ϕ manifests in supercurrents, which are proportional to its gradient. A scanning superconducting quantum interference device (SQUID) susceptometer is highly sensitive to supercurrents [26–29] and therefore suitable for mapping the SC phase. Previous work has demonstrated that SQUID imaging can be used to sense the superfluid density (SFD) [26,30,31]. When phase fluctuations dominate the physics, the SQUID is also sensitive to the phase [32,33].

Here, we use a scanning SQUID to image the spatial evolution of phase coherence in a SC grid and a hybrid JJA. Starting from a continuous SC grid, we show that near $T_c$ the SC continuity in the grid breaks into puddles of varying phase coherence. To better access phase information, we then study a hybrid JJA and tune it with a perpendicular magnetic field. Using an on-chip field coil, we measure the local susceptibility to reconstruct the spatial phase information, even when phase stiffness is extremely small. We detect organized periodic phase patterns at integer and fractional fillings of the unit cell, as previously predicted by theories [34,35]. At a small but finite magnetic field, we map how the JJA breaks into spatially phase-coherent regions where $\xi_C > \xi$. These regions identify specific junctions across which the phases remain coherent. Our measurements provide a high-resolution visualization of phase coherence and its systematic suppression, revealing the long-range phase coherence and phase patterns in JJA.

The paper is organized as follows: we first introduce a continuous grid to demonstrate our sensitivity to SC phase coherence in Section I. In Section II, we observe phase coherence in hybrid JJA, which is maximized at zero magnetic field. In Section III, we reveal an organized checkerboard phase pattern, followed in Section IV by the fragmentation of the phase-coherent region at small but finite magnetic fields. Finally, we conclude by summarizing our results and discussing the prospects for future scanning SQUID measurements on hybrid JJAs.

## I. Continuous Grid

To demonstrate SQUID sensitivity to phase, we measured a continuous Nb SC grid (Fig. 1a). Well below $T_c$, the static magnetic landscape displays quantized flux trapped in the grid cells (Fig. 1b) [36,37]. The filling factor $f$ is defined by the average flux quantum $\Phi_0$ per cell, and is controlled by the external magnetic field present when the grid was cooled into the SC state. At $f = -0.2$, 20% of the cells trap one level (one $\Phi_0$) lower than the rest of the cells (Fig. 1b). These fluxoids are persistent current loops that hold magnetic flux of a single $\Phi_0$. A susceptibility map shows a homogeneous diamagnetic response across the grid (Fig. 1c). The on-chip ac excitation is screened by the SC network, with shielding currents confined to the grid wires, so the susceptibility map reflects the local SC phase stiffness.

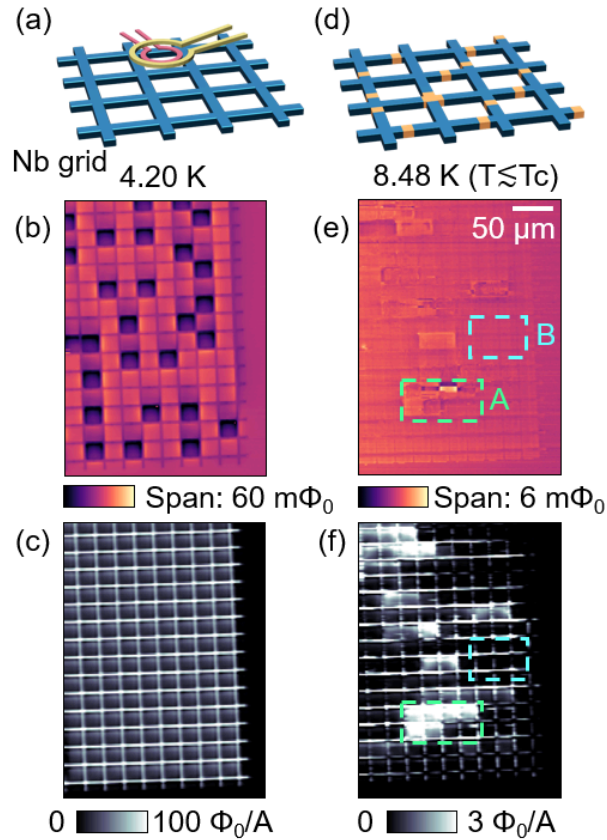

**Fig. 1 Scanning SQUID measurements on a Nb grid.** (a) An illustration of a scanning SQUID microscope, with a fully SC (blue) grid below $T_c$. The lattice constant of the grid is 24 μm with a 4 μm line width. (b) The dc magnetization landscape of the SC grid at 4.2 K, at f = -0.2. The difference between dark purple cells and the rest is one fluxoid (phase winding of 2π). (c) A map of the local susceptibility measured simultaneously with (b), showing the diamagnetic response and a fully uniform SFD. Notice that the supercurrents only flow along the Nb wire-grid. (d) A grid of broken SC continuity by weak links (orange) near $T_c$. (e) Magnetic landscape of the grid near $T_c$ at 8.48 K. Two rectangles mark two different flux behaviors. Region A (green) shows some shielding currents and region B (cyan) shows no shielding currents. (f) Local susceptibility measured simultaneously to (e), showing the diamagnetic response and coherent phase information near $T_c$. The bright loop in region A indicates that the Josephson tunneling dominates, with all junctions in phase coherence and sharing the response to the excitation field. The region B shows only horizontal SC wires, with no closed path to hold persistent current or phase coherence.

Near $T_c$, the SC regions are interrupted by weak links arising from natural inhomogeneities that break the SC continuity in the grid (Fig. 1d). This is reflected by both the static field configuration (Fig. 1e) and the local susceptibility map (Fig. 1f). The signals are 1-2 orders of magnitude weaker, as a result of reduced SFD. Near the transition, weak links act as junctions between SC regions, turning the sample into an irregular JJA. Unlike designed JJA, the weak links or Josephson junctions in this SC grid arise naturally and randomly, reflecting the intrinsic minor disorder of the sample.

In the weakly SC state (Fig. 1e,f), we identify two different regions: region A, where both static flux and local susceptibility show a detectable signal, and region B, where both measurement modes show inactive cells. The difference between the two regions originates from the continuity of the superfluid. Unlike region A, region B does not show trapped flux, since the SC continuity in region B is already broken. When SC grid continuity is broken into non-overlapping regions, the susceptibility image clearly shows active SC wires (region B), but no overall shielding, and correspondingly, weak shielding in the static magnetic image. However, when the SC puddles are connected by Josephson coupling, the image shows both the white SC puddles and also an entire-cell shielding provided by supercurrents revolving around the entire cell (region A). Thus, even with a very weak superfluid stiffness, superconductivity is established across puddles via Josephson coupling and can be visualized by SQUID susceptometry. The dc image shows some flux trapping as well, similar to the low-T image, but with a much weaker signal and visible dynamics. This contrast between regions illustrates how phase coherence is maintained through the coupling across SC puddles. If a closed path of coupled puddles forms, the region can screen with diamagnetic currents flowing around that path, and these puddles will act as a phase coherence cluster.

The diamagnetic reading of the entire cells in region A does not mean that the substrate is SC. It means that there is a continuous loop around multiple grid cells where phase coherence is maintained, and thus, this area screens as a cluster. At every pixel, the SQUID applies the same amount of field and requires a certain amount of screening current to be shielded as a superconductor, thus showing a uniform color in region A. However, if the tunneling is too weak and the region becomes open (region B), the sample is unable to screen the applied field. The Josephson junctions across that region are suppressed and no coherent phase tunneling occurs.

The fact that the susceptibility map (Fig. 1c) is insensitive to the flux trapping (Fig. 1b) well below $T_c$ is a result of a strong diamagnetic response, compared to one fluxoid. Near $T_c$ where the diamagnetic response is much weaker (lower SFD), the phase coherence is highlighted in susceptibility. These measurements demonstrate the high sensitivity of the scanning SQUID, capable of detecting superfluid responses even in weak SC states. By visualizing spatial patterns of phase coherence, we can directly identify the formation of superconductivity at criticality. The similarity between naturally disordered grids and JJAs motivates studies in designed JJAs, where disorder can be controlled, providing a versatile platform to investigate the emergence and breakdown of phase coherence near quantum phase transitions.

## II. Hybrid JJA

To map the phase information in a system with minimal involvement of disorder, we studied an Al/InAs [38,39] hybrid JJA, consisting of an array of SC islands on a square lattice, separated by Josephson junctions (Fig. 2a). The Al islands are in the shape of a cross instead of a square [17–19]. The presence of InAs makes it a unique tunable system in which two sets of gates [19] control both the proximity of the entire two-dimensional electron gas and locally the Josephson tunneling at the junctions. This enables full control over open/closed junctions, as well as high/low carrier density. By tuning both gates, the system can be tuned between the insulating state of fully disconnected islands and the SC state, where large-scale phase coherence exists in the sample. In this work, we keep the gates at zero voltage, resulting in fixed $E_C$ and $E_J$.

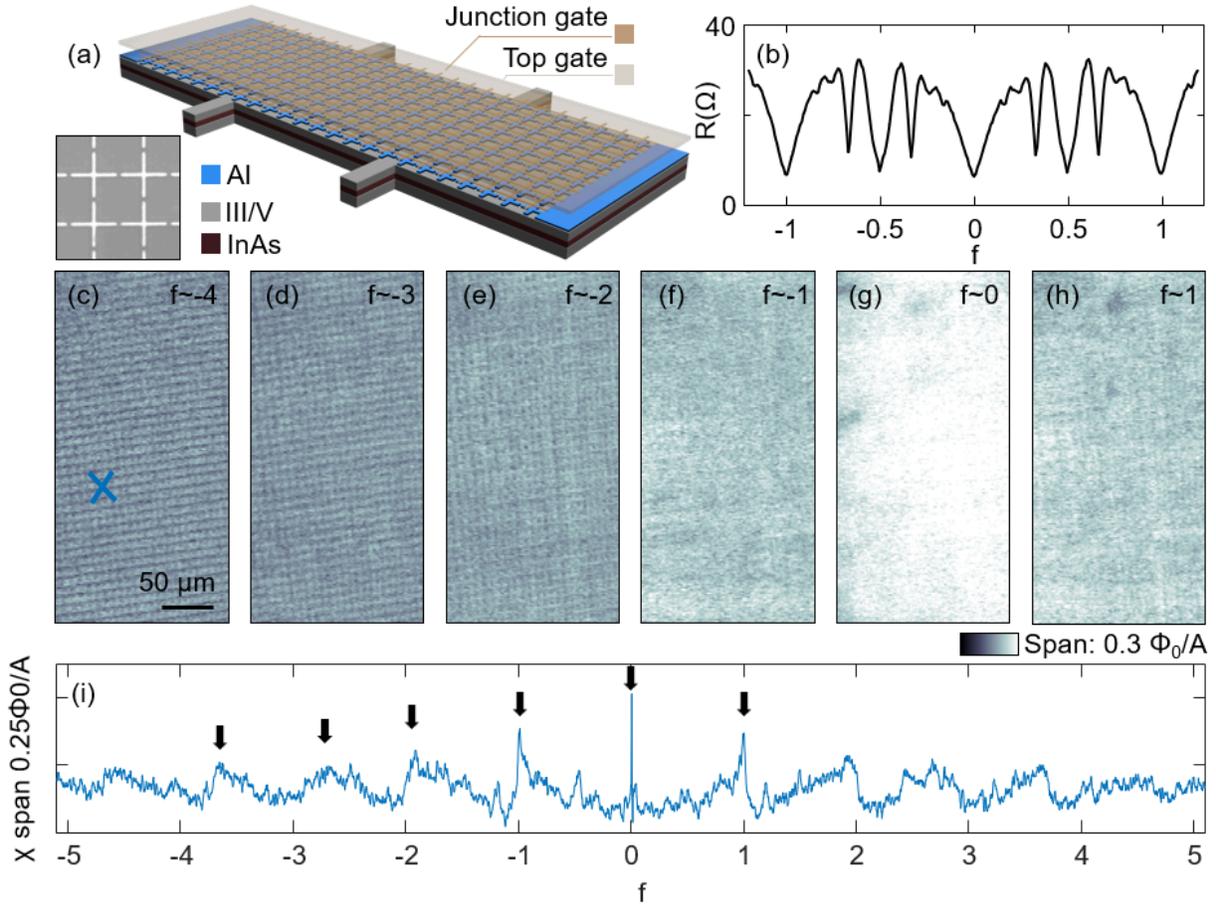

**Fig. 2 Global coherence at zero field and loss of coherence at higher integer filling.** (a) Illustration of the layers in the hybrid JJA. The left inset shows a scanning electron microscope image of the Al crosses. The lattice constant of the crosses used in this paper is 3 μm, and the linewidth is 0.3 μm. (b) Resistance oscillation in JJA at T = 20 mK, I = 1.4 μA, with a period of $\Phi_0$. The resistance reaches global minima at integer fillings and local minima at fractional fillings. Dips at 1/2 and 1/3 are prominent, while there are less prominent dips at 1/4 and 1/5. (c–h) Diamagnetic response of the JJA at integer fillings from −4 to +1, at 20 mK. Near the zero field (g), the diamagnetic response is the strongest, and no grid structure is seen, resembling full phase coherence in a superconductor. At higher integer filling, the phase coherence is gradually lost and the grid structure is revealed. The black arrows in (i) mark the locations where (c-h) were measured. (i) Susceptibility measurement of a single point as a function of the field. The location of this point is marked by a blue cross in panel (c). The height of the peaks gradually diminishes while remaining symmetric compared to f = 0, indicating the decrease in phase coherence at higher integer filling.

At a temperature well below $T_c$, the crosses are SC, but unlike a continuous grid, the SC islands communicate mainly via Josephson tunneling. The phase coherence of this system can be tuned by applying a global magnetic field, which induces phase gradients into the system. When the magnetic field changes, the system exhibits resistance oscillations [40]. We measured the longitudinal resistance $R$ as a function of filling fraction $f$, far below $T_c$. The results are shown in Fig. 2b, where dips in $R$ are observed at integer and fractional $f$, resembling the Little-Parks effect [41]. Unlike the continuous SC grids, which exhibit resistance and global susceptibility oscillations only close to $T_c$ [42–44], a JJA exhibits resistance oscillations well below $T_c$ [18,45]. So, for JJA, we expect to find phase-related patterns in the susceptibility signal already at base T.

We mapped the spatial distribution of the local susceptibility as a function of the global magnetic field. We used the resistance oscillation as a guide to explore the phase coherence. Fig. 2g shows a spatial map of the local susceptibility at $f = 0$. Note that the measured signal is yet another order of magnitude lower than in Fig. 1f, and that no signal is observed in the dc flux. The measured signal of 0.2 $\Phi_0$/A reflects a weak diamagnetic response. For comparison, at the same (base) temperature, we detected a diamagnetic response of 80 $\Phi_0$/A near an Al pad on the same sample. However, given the small width of the crosses and the porous geometry of JJA, the diamagnetic response from the JJA is weaker and comparable to other materials closer to 2D behavior, where the Pearl penetration depth is in the millimeter range (see Appendix).

The delicate diamagnetic response from the JJA can also be easily suppressed by the local magnetic excitation used to generate the susceptibility map, which further emphasizes the sensitivity of local susceptometry to fragile SC states. At $f = 0$ the diamagnetic response is relatively homogeneous, indicating that the entire sample is phase coherent and screens together under our local magnetic excitation.

Local susceptibility maps at increasing integer $f$ values show that the measured signal decreased (Figs. 2c–h), indicating suppression of phase stiffness and the SC strength. As the spatial maps are relatively homogeneous, we measured the local susceptibility at a fixed location in the sample while $f$ continuously sweeping over a large range (see Supplemental Material Fig. S1). The results reveal clear peaks of decreasing amplitude at integer $f$, along with a sawtooth-like background (Fig. 2i). Note that as $f$ increases, the effective unit cell area grows due to weakening of the proximity effect, explaining the shift of the peaks from the expected integer filling values. This decrease in phase coherence is consistent with the transport result in a similar hybrid JJA [46].

As $f$ is tuned between consecutive integers, the susceptibility exhibits a characteristic sawtooth-like form (Fig. 2i). At integer filling, the vortex lattice is commensurate with the underlying grid, leaving no mobile vortices. Phase fluctuations are therefore minimal, phase stiffness is maximal, and the susceptibility peaks. Slightly away from commensurability, a small number of mobile vortices are introduced. Their ability to arrange in many configurations increases the entropy but reduces the stiffness, causing a sharp drop in susceptibility. Close to commensurate fillings, interactions constrain the possible arrangements, partial ordering emerges, and the stiffness gradually recovers, culminating in the next peak at $f = n + 1$. At higher integers, the peak amplitude decreases, since the SC phase around each grid cell must twist more to accommodate multiple vortices.

### III. Phase patterns at fractional $f$

The amplitude of the local susceptibility already reveals rich physical insights. Yet, the spatial phase configuration holds further information, which becomes evident at fractional values of $f$. At $f = 0$ the response is spatially homogeneous, indicating that the entire array is phase coherent and screens collectively (Fig. 3a). The phases of every cross are aligned (Fig. 3b) and evolve together. When tuned to an incommensurate filling, the global coherence is strongly interrupted: the individual SC crosses remain locally SC, but coherence across the junctions is diminished, and flux within each grid cell is not efficiently screened (Fig. 3c). The corresponding phase pattern shows sparse phase vortices (Fig. 3d), which are unstable under the local excitation field. In the vortex-propagation picture, this corresponds to incommensurate vortices that propagate freely through the array, generating local phase fluctuations and weakening the phase stiffness. In Fig. 3e, we show the 3 µm array structure of Fig. 3c using the fast Fourier transform (FFT).

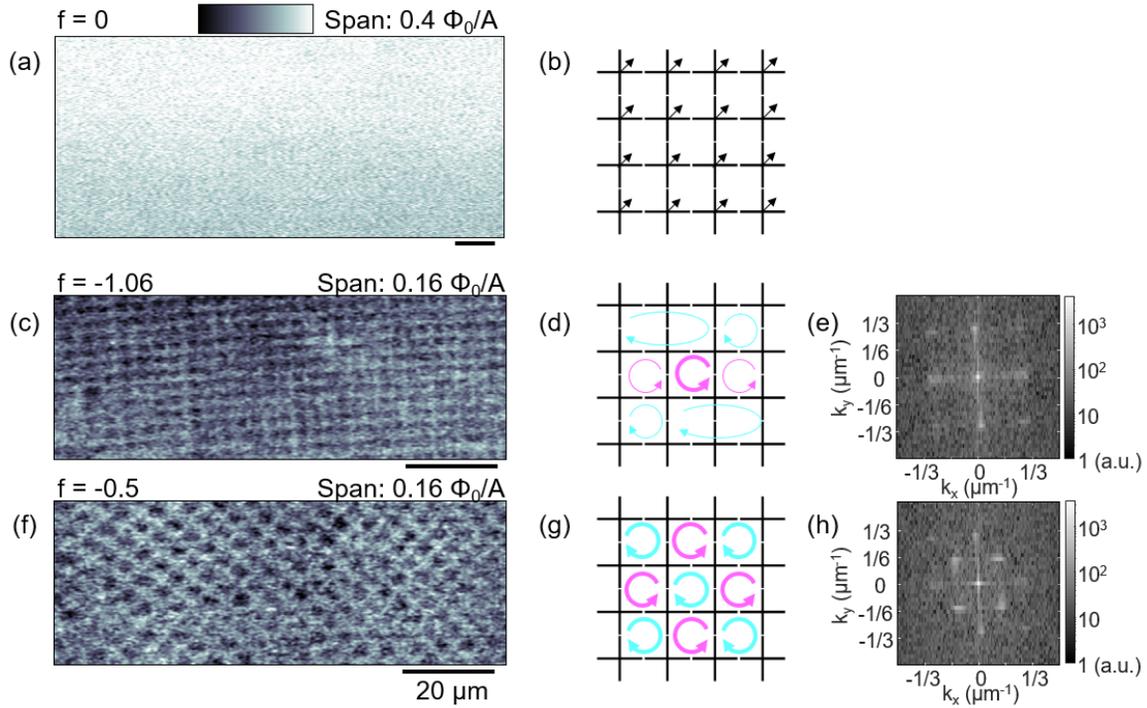

**Fig. 3 Phase patterns at commensurate and incommensurate fillings.** (a) Diamagnetic response at f = 0, showing a homogeneous SC texture across the JJA. (b) Corresponding phase map: the phases of all SC islands are coherent and aligned to minimize the total energy. (c) At an incommensurate filling (f = –1.06), the susceptibility map reveals the underlying grid structure. (d) Schematic phase pattern of (c), with circles indicating Josephson current directions. Incommensurability introduces complex phase textures and suppresses global coherence. (e) FFT of (c), highlighting the grid periodicity at 1/3 µm−1. (f) At a commensurate filling of –0.5, the susceptibility map exhibits a checkerboard pattern. (g) The schematic phase map of (f) illustrates alternating clockwise and counterclockwise Josephson currents. (h) FFT of (f), showing the checkerboard frequency at 1/6 µm−1, corresponding to a doubled real-space period. All measurements are performed at 20 mK.

At the commensurate fractional filling f = −1/2, the susceptibility develops a checkerboard pattern (Fig. 3f). The FFT (Fig. 3h) reveals peaks at half the reciprocal lattice vector, consistent with a doubled spatial period compared to the grid structure at an incommensurate filling (Fig. 3e). The corresponding phase map shows a pattern rotated by 45° relative to the grid, reflecting a checkerboard pattern with an average of half a flux quantum per cell. Each cell compensates by screening currents that either subtract or add half a flux quantum, yielding alternating clockwise and counterclockwise currents (Fig. 3g). This produces two degenerate checkerboard states, and stochastic switching between them was observed during scanning (Fig. S2). Similar higher-order periodic states appear at another fractional filling 1/3 (Fig. S5). To our knowledge, this is the first direct visualization of such fractional phase patterns in JJA.

## IV. Phase-coherent regions around $f = 0$

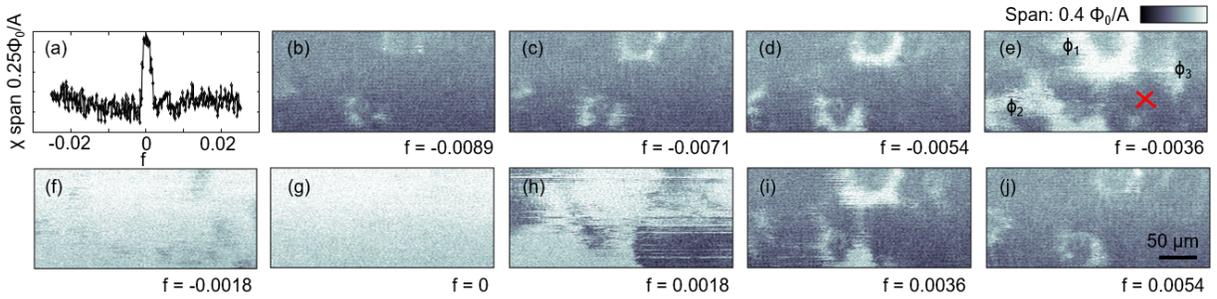

Fig. 4. Phase-coherent regions near f=0 in susceptibility maps. (a) Susceptibility of a single point as a function of the field. The location of this point is marked by a red cross in panel (e). When the regions expand and shrink through this point, the diamagnetic response peaks. (b-j) The evolution of the phase-coherent regions as a function of the field. As the field is swept away from zero, the full region (g) breaks into smaller parts (e). Between each image, the field was increased by 0.4 µT. In panel (e), all elements inside the same region share the same phase and are independent of the phases of other regions. Within each region, all elements remain phase coherent. All measurements are performed at 20 mK.

At small but finite fields, the extended phase coherence forms and breaks over a narrow region (Fig. 4). We spatially mapped the evolution of the phase-coherent regions. The entire phase-coherent region diminishes quickly with the applied field. At zero field (Fig. 4g), the entire scan range displays a homogeneous diamagnetic response, suggesting that the phase-coherence length scale at zero field is larger than the image size ($\xi_C$ > 200 µm) and possibly extends over the entire sample to the millimeter scale. When we slightly shifted the magnetic field away from zero (Figs. 4f, h), we observed fluctuating behavior that disrupted the coherent picture. Furthermore, the sample breaks into smaller, isolated regions (Figs. 4e, i). This happens due to the interruption of phase coherence by the field. The fact that different regions show the same diamagnetic response suggests that the JJA has a sufficient amount of superfluid to repel the applied field in those regions completely. Each SC island (a cross) within a region remains phase-coherent with all other crosses in the same region, and collectively screens the applied field. The individual regions become phase independent and disconnected from other regions. To interpret this picture, each region can be assigned a phase that evolves independently over time ($\phi_1$, $\phi_2$ and $\phi_3$ in Fig. 4e). Sweeping $f$ away from zero by 0.01, the local phase coherence is lost and the grid structure is revealed (Fig. 4b). This means that every cross has very limited phase coherence with its neighbors, as shown in Fig. 3c. These phase-coherent regions directly visualize the dip in transport measurements near zero field [17,18]. The suppression of long-range phase coherence by infinitesimally small frustration may explain the previously observed lifting of the resistivity saturation in the context of the anomalous metallic phase [19].

Specific shapes of the regions, as seen in Fig. 4, tend to repeat. This is not a surprise, because in virtually all samples near the transition, superconductivity always breaks down inhomogeneously, revealing even the most delicate modulations in SC properties determined by $T_c$ variations and local defects [47]. We have cycled the sample temperature to 4 K, well above the $T_c$ of the sample, and found that the overall pattern of the regions repeats, but the details of the regions, especially at the edges, may change (see Fig. S6). The pattern also remains stable over field sweep repetitions (Fig. S7), or hours of waiting time (Fig. S8).

At the center of the phase-coherent regions, incoherent normal areas sometimes appear. We interpret these holes (see Fig. 4e and Fig. S9) as phase vortices. To see how much flux each phase-vortex carries, we count the number of cells in each, which requires precise knowledge of the superfluid map, and calculate the area of non-SC regions. A rough estimate of the flux in each gives an order of one $\Phi_0$ (see Fig. S9), supporting our hypothesis. This is a clear demonstration of how JJAs simulate the behavior of a superconductor. The number of normal cores increases with the applied magnetic field, and some cores appear or disappear, similar to the behavior of vortices.

## V. Discussion and summary

To summarize, we have used scanning SQUID susceptometry to capture both the amplitude and phase information of the order parameter. In hybrid JJA, we use a global magnetic field to tune the phase coherence between SC crosses and extract phase information from the magnetic susceptibility map. At zero field, each element in the JJA is long-range phase-coherent, possibly extending throughout the entire device. The phase coherence leads to a diamagnetic response which peaks at integer fillings and decays under higher magnetic fields. We also detect periodic phase patterns at fractional commensurate fillings 1/2 and 1/3. These phase patterns confirm that phase coherence can be imaged even in the case of weak superconductivity, with signals three orders of magnitude weaker than those of a fully SC Nb film. Furthermore, we observe that the phase-coherent region expands and contracts rapidly in small but finite fields. This highlights the ability of scanning SQUID to map sensitive phase information over long ranges.

The ability to directly image local susceptibility with scanning SQUID opens many directions. Hybrid JJAs can be tuned between SC and insulating ground state by controlling electron density and inter-island coupling [19], and they can be used to access open questions in condensed matter physics, such as the mechanism of the anomalous phase [17,19,48] near the SIT. Another potential local examination is to trigger phase slips in JJA with the SQUID field coil, similar to current-phase relations (CPR) measurement [29,49]. This provides deeper insight into the vortex interactions and lattice-dependent frustration.

Beyond gate control, hybrid JJAs provide a versatile platform that can be engineered with complex lattice geometries [50] (such as triangular, kagome, or dice lattices [46,51]), different junction geometries [52], or π-junctions. With appropriate geometries, JJAs can host unconventional phases [51] and function as topologically protected qubits [53–55]. JJAs can also be designed to introduce artificial dislocations and positional disorder [56–59]. Recent theoretical work further predicts that precisely controlled local phase can drive JJAs into a topological SC regime [52,60].

Together, these perspectives highlight JJAs as a highly tunable model system and establish scanning SQUID susceptometry as a uniquely powerful probe of their fragile phase coherence.

**Acknowledgments**

We thank I. Volotsenko and A. Frydman for fabricating the Nb grid sample, and S. Kallatt for fabricating the hybrid JJA. X.W. and B.K. were supported by the European Research Council No. ERC COG 866236, the ISF, No. ISF-228/22, German-Israeli Project Cooperation DIP No. KA 3970/1-1, and the COST Action CA21144. A.C.C.D, C.M.M., and S.V. acknowledge support from research grants (Projects No. 43951 and No. 53097) from VILLUM FONDEN, the Danish National Research Foundation, and the European Research Council (Grant Agreement No. 856526). N.T. acknowledges support of the National Science Foundation through award DMR-2138905.

**Appendix: Methods**

    I.    **Sample fabrication**

The continuous Nb grid sample is 50-nm-thick, deposited on a silicon substrate by e-beam evaporation under UHV conditions. The grid was patterned using standard photolithography and lift-off with a Heidelberg Instruments MLA 150 system. The sample was used in Ref. [47].

The hybrid JJA was fabricated using standard e-beam lithography techniques on a shallow InAs quantum well heterostructure grown on InP and terminated with 8 nm epitaxial Al. The same heterostructure was used in Ref. [39]. First, a mesa was formed, etching 300 nm into the heterostructure. Al was stripped everywhere except the active regions on the mesa, and ohmic contacts for bonding were made by evaporating Ti/Au. Alignment marks were deposited before a high-precision Transene-D Al etch of the JJA on the mesa. A first ALD of 15 nm hafnia was deposited, followed by a 3/16 nm Ti/Au junction gate grid aligned to the JJA. Electrical connection to the junction gate was realized by a stack of 10/390 nm Ti/Au, climbing up the mesa and connecting to a bonding pad. A second ALD of 15 nm hafnia was deposited, followed by 5/40 nm Ti/Au top gate covering the mesa and another 10/390 nm Ti/Au deposition for contacting to a bond pad. In the measured sample, both gates were shorted and therefore could not be used in this study. The gates were kept at a fixed electrostatic potential of 0 V throughout the experiment.

    II.    **Scanning SQUID susceptometry**

We use a planar SQUID to scan the SC landscape of the sample [27,28]. The SQUID sensor used in the grid sample is composed of a concentric 1.5 µm diameter pickup loop and an 8 µm field coil. For studying the JJA, we use a pickup loop 1.5 µm and a field coil 6 µm. We use the pickup loop to measure the static magnetic flux, which is shown in the Figs. 1b, e. The local ac susceptibility is also recorded, simultaneously with the static magnetic landscape by applying an ac current through the local pickup loop (Fig. 1a) at typical frequencies in the order of kilo-Hz. The flux through the SQUID at that frequency measures the ac diamagnetic response equivalent to the susceptibility (Figs. 1c, f).

As the SQUID dimensions are smaller than the grid cell, strong peaks in susceptibility (white) are observed when the SQUID is directly above the grid wires, as local currents in the SC grid wires screen the local field. By contrast, when the SQUID is above the empty grid cell, a washed-out peak is observed in susceptibility as the screening currents flow farther from the SQUID around the perimeter of the grid cell, and part of the excitation field lines close before reaching the grid wires. This homogeneous pattern across the entire grid directly reproduces the physical dimensions of the grid.

The amplitude of the susceptibility measurement contains information about the penetration depth. A susceptibility of 180 $\Phi_0$/A was detected for a 50-nm-thick Nb film below $T_c$ [47], corresponding to a penetration depth of 47 nm [61,62]. Naturally, the geometry of the sample and sensor should be taken into account when comparing these mutual inductance measurements [63]. For example, in Fig. 1c, the porous 50-nm-thick Nb grid only generates 100 $\Phi_0$/A. The SQUID can also sense the amplitude of the phase stiffness in a relatively weak SC state. Previous measurements estimated the Pearl penetration depth of 8 mm at the LaAlO$_3$/SrTiO$_3$ interface, corresponding to a diamagnetic response of 0.4 $\Phi_0$/A; and 0.65 mm in $\delta$-SrTiO$_3$, with a response of 4 $\Phi_0$/A [30,31].

# Supplemental Material

## I. Autocorrelation analysis

To characterize the phase coherence quantitatively, we perform autocorrelation analysis of the susceptibility map. First, we extract the susceptibility data S(x,y) (Fig. S1a) and calculate its autocorrelation S'(x,y) (Fig. S1b). Next, we transform the x-y axes into polar coordinates (r–θ) and average over θ. At this stage, all the anisotropic information is left out, such as the shape of the phase-coherent regions, but focusing on the strength of the regions. The averaged autocorrelation function S'(r) is then fitted to an exponential decay model $Ae^{-r/\lambda}$, where A represents the diamagnetic strength of the regions, and λ the decaying length scale (Fig. S1c). We mainly focus on the result of amplitude A, because the estimated length scale λ is limited by the scan range, whereas the main text Fig. 4 already indicates device-wide correlation.

Fig. S1d shows the calculated A as a function of the filling factor, revealing a sharp peak near zero and additional peaks and correlations near commensurate and integer filling. The peak locations match the SQUID susceptibility measurement at a single point in the main text Fig. 4 (Fig. S1e), indicating enhanced correlation near commensurate filling. Notably, the period obtained from transport measurement (Fig. S1f) is slightly different from the period detected locally by following the evolution of the phase-coherent region. This discrepancy may arise from the difference between the physical area of the cells and the actual landscape of SC, which occupies some of the cell area.

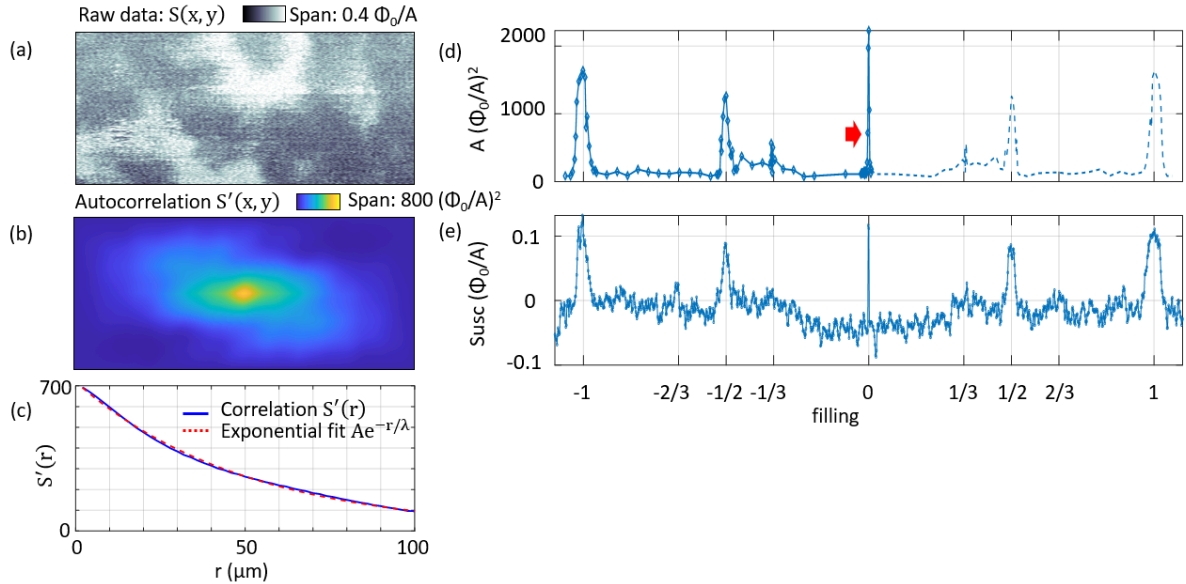

**Fig. S1. Analysis of the correlation strength.** (a) Susceptibility raw data S(x,y), representing the SFD map at f = −0.0036. The size of this image is 223 μm * 100 μm. (b) The autocorrelation map S'(x,y) of the panel a. It was calculated according to the 2D cross-correlation of S(x,y) with itself. (c) Exponential fit of the autocorrelation S'(r), where the parameter A=720 $(\Phi_0/A)^2$ represents the strength of the phase-coherent regions and λ = 49

µm is the decaying length scale. (d) The strength of region A as a function of the filling factor. Each point represents the result of a correlation analysis from a single scan. Data was collected for filling factors in the range between –1 and 0.01. For f>0.01, the data is duplicated and flipped (dashed line). The red arrow marks the data of panel a. (e) Susceptibility measurement at the same location marked by the red cross in Fig. 4e (the same location as Fig. 4a), as a function of the field from f =–1.05 to 1.05. All measurements are performed at 20 mK.

## II. Two degenerate checkerboard patterns

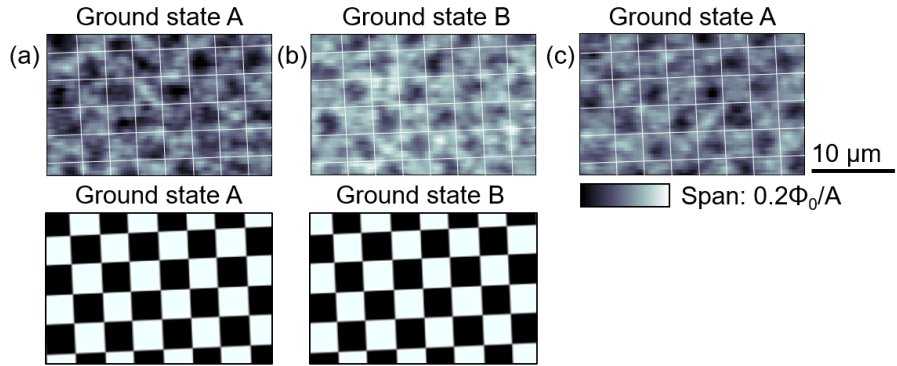

**Fig. S2. Susceptibility maps of checkerboard phase pattern at f = –0.5.** The sample switches between two degenerate ground states in successive scans, at 20 mK. To better demonstrate the periodicity of the checkerboard, we add the auxiliary white lines representing the 3 µm grid structure.

## III. Fast Fourier Transform analysis

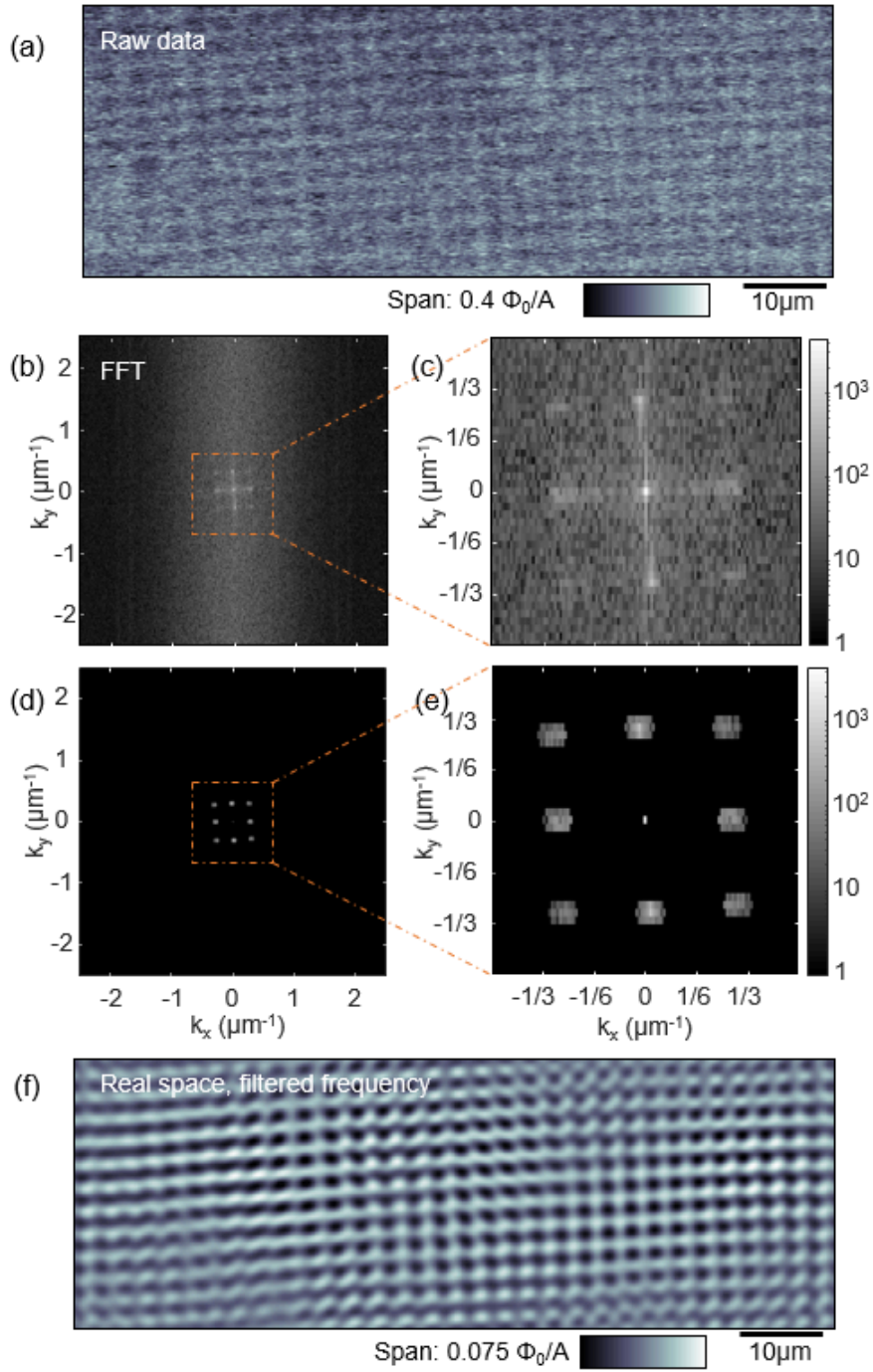

Fig. S3. FFT to show the lattice constant of the sample. (a) Raw data of susceptibility at f = −1.06, the same data as shown in the main text Fig. 3c. (b) FFT of the raw data and (c) the zoom-in area to highlight multiple prominent frequencies. (d-e) Applying a filter only to select the frequencies near 1/3 µm^−1, which are relevant for the 3 µm lattice constant. (f) After transforming back to the real space, a clear pattern of grid structure is shown.

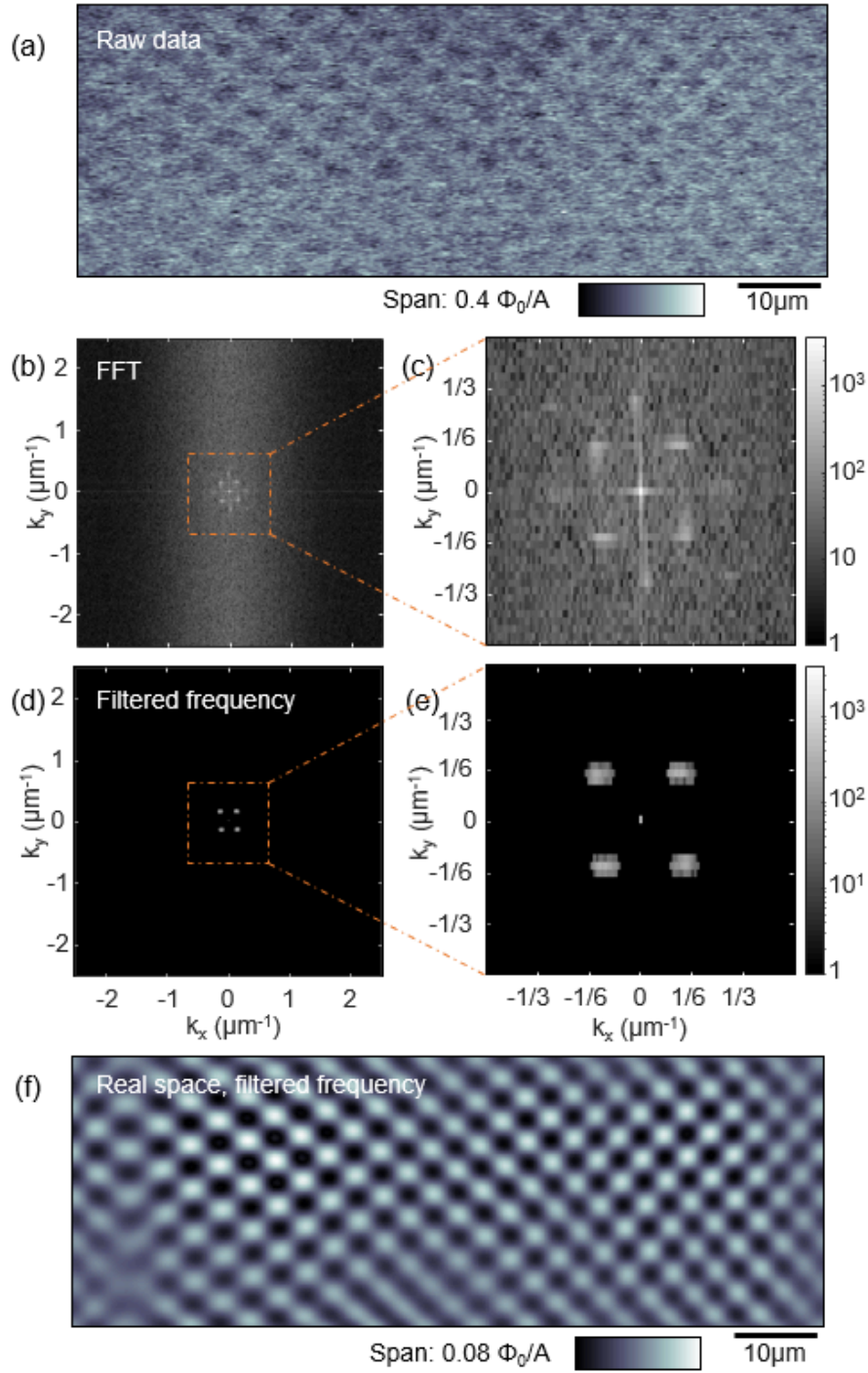

**Fig. S4. FFT to prove the phase pattern of the checkerboard.** (a) Raw data of susceptibility at f = −0.5, the same data as shown in the main text Fig. 3f. (b) FFT of the raw data and (c) the zoom-in area to highlight multiple prominent frequencies. (d-e) Applying a filter only to select the frequencies near 1/6 µm^-1, which are relevant for a checkerboard (two times the lattice constant). (f) After transforming back to the real space, a clear checkerboard pattern is revealed.

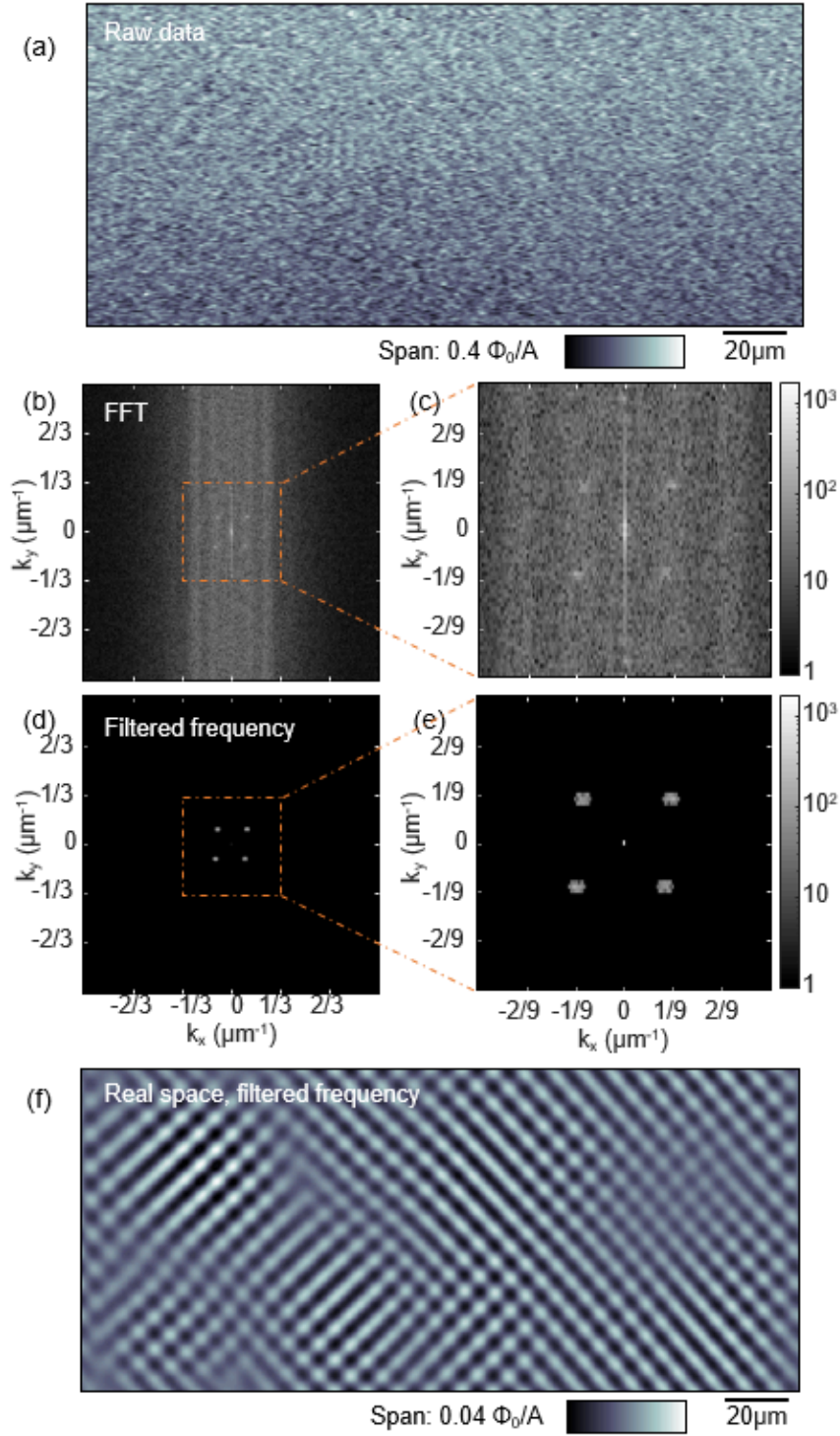

**Fig. S5. FFT to prove the phase pattern of the staircase, theoretically expected for f = 1/3.** (a) Raw data of susceptibility at f = −0.33. (b) FFT of the raw data and (c) the zoom-in area to highlight multiple prominent frequencies. (d–e) Applying a filter only to select the frequencies near 1/9 µmˆ-1, which are relevant for a staircase pattern (three times the lattice constant). (f) After transforming back to the real space, a clear pattern of diagonal lines is revealed. The prominent periodicity of 9 µm proves that these lines are the staircase pattern.

## IV. Tests regarding sample inhomogeneity

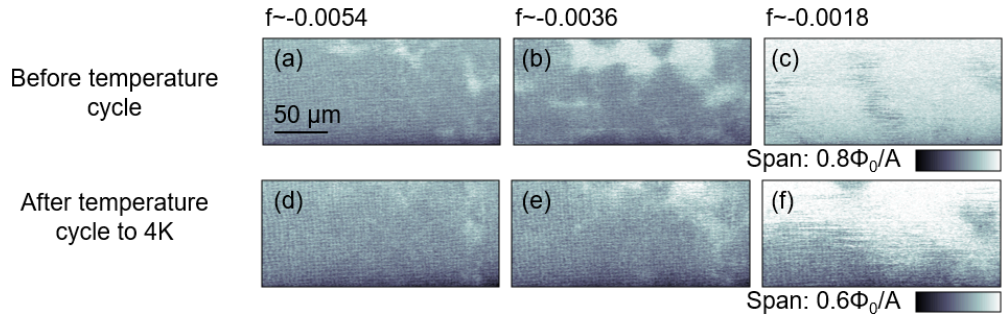

**Fig. S6. Susceptibility maps of phase-coherent regions as a function of the magnetic field before (a–c) and after (d–f) the temperature cycle to 4 K.** The shape remains mostly unchanged upon cycling the temperature to 4 K (above $T_c$ of thin aluminum, 1.3 K) and then back to the base temperature around 20 mK.

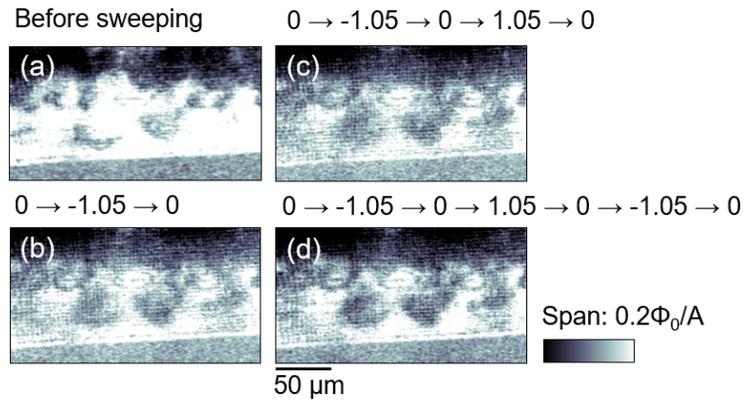

**Fig. S7. Susceptibility maps of phase-coherent regions near zero field, with different field sweeping histories.** The shape barely changes upon sweeping the magnetic field, and then back to the same field. All measurements are performed at 20 mK.

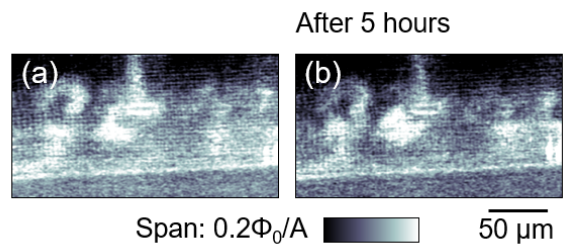

**Fig. S8. Susceptibility maps of phase-coherent regions near zero field at 20 mK, after waiting for a long time.** The shape barely changes after 5 hours.

## V. Phase vortices formation

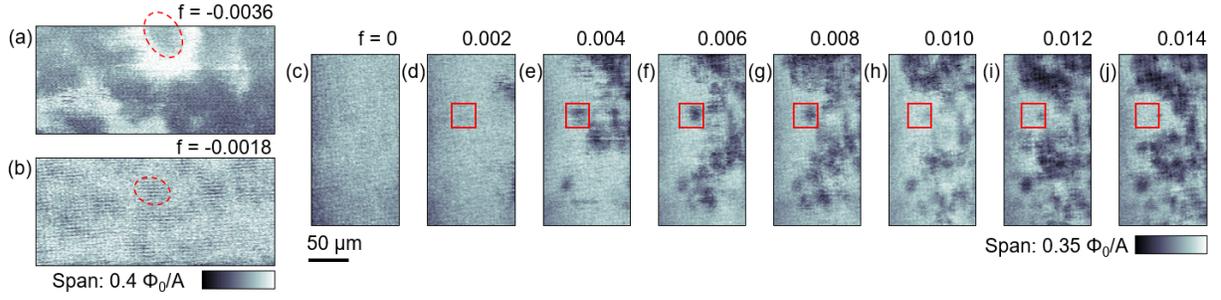

**Fig. S9. Susceptibility maps of phase-coherent regions near zero field, at different areas of the sample.** We observed normal areas located at the center of phase-coherent regions (marked by red circles), resembling phase vortices. The flux penetrating these cores in the red ovals is estimated to be (a) 0.48 $\Phi_0$, (b) 0.23 $\Phi_0$, in the order of one $\Phi_0$. We estimate this by counting (1) the total number of cells in the scan, $N$; (2) the number of phase-coherent cells, $N_c$; and (3) the number of cells inside the hole, $N_h$. The field penetrating the hole is then $fNN_h/(N - N_c)$ where $f$ is flux per unit cell. (c–j) The number of individual normal areas grows with the field. The red square marks an area where the black phase vortex appears, but not always, resembling the behavior of a real vortex. All measurements are performed at 20 mK.